\documentclass[12pt]{article}
\usepackage{a4wide}
\usepackage{epsfig}
\usepackage{amsmath}

\newlength{\absize}
\catcode`@=11
\def\citer{\@ifnextchar [{\@tempswatrue\@citexr}{\@tempswafalse\@citexr[]}}

\def\@citexr[#1]#2{\if@filesw\immediate
  \write\@auxout{\string\citation{#2}}\fi
  \def\@citea{}\@cite{\@for\@citeb:=#2\do
    {\@citea\def\@citea{--\penalty\@m}\@ifundefined
       {b@\@citeb}{{\bf ?}\@warning
       {Citation `\@citeb' on page \thepage \space undefined}}%
\hbox{\csname b@\@citeb\endcsname}}}{#1}}
\catcode`@=12

\begin{document}
  \thispagestyle{empty}
  \pagestyle{empty}
  \renewcommand{\thefootnote}{\fnsymbol{footnote}}
\newpage\normalsize
    \pagestyle{plain}
    \setlength{\baselineskip}{4ex}\par
    \setcounter{footnote}{0}
    \renewcommand{\thefootnote}{\arabic{footnote}}
\newcommand{\preprint}[1]{%
  \begin{flushright}
    \setlength{\baselineskip}{3ex} #1
  \end{flushright}}
\renewcommand{\title}[1]{%
  \begin{center}
    \LARGE #1
  \end{center}\par}
\renewcommand{\author}[1]{%
  \vspace{2ex}
  {\Large
   \begin{center}
     \setlength{\baselineskip}{3ex} #1 \par
   \end{center}}}
\renewcommand{\thanks}[1]{\footnote{#1}}
\begin{flushright}
\end{flushright}
\vskip 0.5cm

\begin{center}
{\large \bf Un-equivalency Theorem of Deformed Heisenberg-Weyl's
Algebra in Noncommutative Space}
\end{center}
\vspace{1cm}
\begin{center}
Jian-Zu Zhang
\end{center}
\vspace{1cm}
\begin{center}
Institute for Theoretical Physics, East China University of
Science and Technology, \\
Box 316, Shanghai 200237, P. R. China
\end{center}
\vspace{1cm}
\begin{abstract}
An extensively tacit understandings of equivalency between the
deformed Heisenberg-Weyl algebra in noncommutative space and the
undeformed Heisenberg-Weyl algebra in commutative space is
elucidated. Equivalency conditions between two algebras are
clarified. It is explored that the deformed algebra related to the
undeformed one by a non-orthogonal similarity transformation.
Furthermore, non-existence of a unitary similarity transformation
which transforms the deformed algebra to the undeformed one is
demonstrated. The un-equivalency theorem between the deformed and
the undeformed algebras is fully proved. Elucidation of this
un-equivalency theorem has basic meaning both in theory and
practice.
\end{abstract}


\clearpage

Spatial noncommutativity is an attractive basic idea for a long
time. Recent interest on this subject is motivated by studies of
the low energy effective theory of D-brane with a nonzero Ns - NS
$B$ field background \citer{DH,DN}. It shows that such low energy
effective theory lives on noncommutative space. For understanding
low energy phenomenological events quantum mechanics in
noncommutative space (NCQM) is an appropriate framework. NCQM have
been extensively studied and applied to broad fields
\citer{CST,JZZ04b}. But up to now it is not fully understood.

In literature there is an extensively tacit understandings about
equivalency between the deformed Heisenberg-Weyl algebra in
noncommutative space and the undeformed Heisenberg-Weyl algebra in
commutative space. As is well known that the deformed phase space
variables are related to the undeformed ones by a linear
transformation. This leads to such tacit understandings of
equivalency between two algebras.

In this paper we elucidate this subtle point. First we clarify
equivalency conditions between two algebras. Then we demonstrate
that the deformed algebra is related to the undeformed one by a
similarity transformation with a non-orthogonal real matrix.
Furthermore, we prove that a unitary similarity transformation
which transforms two algebras to each other does not exist. The
results are summarized in the un-equivalency theorem between two
algebras. Because the deformed and undeformed Heisenberg-Weyl
algebras are, respectively, the foundations of noncommutative and
commutative quantum theories, elucidation of the un-equivalency
theorem has significant meaning both in theory and practice.
One expects that essentially new effects of spatial
noncommutativity may emerge from noncommutative quantum theories.
This depends on the fact that the deformed Heisenberg-Weyl algebra
is not equivalent to the undeformed one. The un-equivalency
theorem shows that explorations of essentially new effects of
spatial noncommutativity emerged from noncommutative quantum
theories can be expected.

In order to develop the NCQM formulation we need to specify the
phase space and the Hilbert space on which operators act. The
Hilbert space is consistently taken to be exactly the same as the
Hilbert space of the corresponding commutative system \citer{CST}.
As for the phase space we consider both position-position
noncommutativity (position-time noncommutativity is not
considered) and momentum-momentum noncommutativity
\cite{DN,JZZ04a}. In this case the consistent deformed
Heisenberg-Weyl algebra is as follows:
\begin{equation}
\label{Eq:xp}
[\hat x_{i},\hat x_{j}]=i\xi^2\theta\epsilon_{ij},
\qquad [\hat x_{i},\hat p_{j}]=i\hbar\delta_{ij}, \qquad
[\hat p_{i},\hat p_{j}]=i\xi^2\eta\epsilon_{ij},\;(i,j=1,2),
\end{equation}
where $\theta$ and $\eta$  are the constant, frame-independent
parameters.
Here we consider the intrinsic momentum-momentum noncommutativity.
It means that the parameter $\eta$, like the parameter $\theta$,
should be extremely small.
$\epsilon_{ij}$ is an antisymmetric unit tensor,
$\epsilon_{12}=-\epsilon_{21}=1,$ $\epsilon_{11}=\epsilon_{22}=0$;
$\xi=(1+\theta\eta/4\hbar^2)^{-1/2}$ is the scaling factor. It
plays essential role for guaranteeing consistency of the framework
\citer{JZZ04a}.
When $\eta=0,$ we have $\xi=1$. The the deformed Heisenberg-Weyl
algebra (\ref{Eq:xp}) reduces to the one of only position-position
noncommuting.

The deformed phase space variables $\hat x_{i}$ and $\hat p_{i}$
are related to the undeformed ones $x_{i}$ and $p_{i}$ by the
following linear transformation \cite{JZZ04a}
\begin{equation}
\label{Eq:hat-x-p}
\hat x_{i}=\xi(x_{i}-\frac{1}{2\hbar}\theta\epsilon_{ij}p_{j}),
\quad
\hat p_{i}=\xi(p_{i}+\frac{1}{2\hbar}\eta\epsilon_{ij}x_{j}).
\end{equation}
where $x_{i}$ and $p_{i}$ satisfy the undeformed Heisenberg-Weyl
algebra
\begin{equation}
\label{Eq:xp1}
[x_{i},x_{j}]=[p_{i},p_{j}]=0, [x_{i},p_{j}]=i\hbar\delta_{ij}.
\end{equation}

In literature the point of the tacit understandings of equivalency
between the deformed Heisenberg-Weyl algebra and the undeformed
one is as follows: any Lie algebra generated by relations
$[X_a,X_b]=iT_{ab}$ with central $T_{ab}$ satisfying
$det(T_{ab})\ne 0$ can be put into a usual canonical form, like
Eqs.~(\ref{Eq:hat-x-p}). Therefore the deformed Heisenberg-Weyl
algebra (\ref{Eq:xp}) and the undeformed one (\ref{Eq:xp1}) are
the same. Furthermore, the spectrum of an observable is the same
regardless we star with deformed variables $(\hat x_{i},\hat
p_{i})$ or undeformed ones $(x_{i},p_{i})$.

Now we elucidate this subtle point. Equivalency between the
deformed and the undeformed Heisenberg-Weyl algebra must satisfy
two conditions: (i) Two sets of phase space variables $(\hat
x_{i}, \hat p_{i})$ and $(x_{i}, p_{i})$ can be related to each
other by a singular-free linear transformation (The inverse
transformation should exit for all values of $(\hat x_{i}, \hat
p_{i})$ and $(x_{i}, p_{i})$);
(ii) Two algebras can be transformed to each other by a unitary
similarity transformation.

First we consider the second condition. We prove the following
theorem.

{\bf The Un-equivalency Theorem} {\it The deformed Heisenberg-Weyl
algebra in noncommutative space is transformed to the undeformed
Heisenberg-Weyl algebra in commutative space by a similarity
transformation with a non-orthogonal real matrix. A unitary
similarity transformation which relates two algebras to each other
does not exist.}

We define a $1\times 4$ column matrix $\hat U=(\hat U_1,\hat
U_2,\hat U_3,\hat U_4)$ with elements $\hat U_1=\hat x_1$, $\hat
U_2=\hat x_2$, $\hat U_3=\hat p_1$ and $\hat U_4=\hat p_2$, a
$4\times 1$ row matrix $\hat U^T$ with elements $\hat U_i^T=\hat
U_i$, $(i=1,2,3,4)$, and a $4\times 4$ matrix $\hat M$ with
elements
\begin{equation}
\label{Eq:hat-Mij}
i\hat M_{ij}=[\hat U_i, \hat U_j^T], (i, j=1,2,3,4).
\end{equation}
The matrix $\hat M$ represents the deformed Heisenberg-Weyl
algebra. From Eqs.~(\ref{Eq:xp}) and (\ref{Eq:hat-Mij}) it follows
that $\hat M$ reads
\begin{equation}
\label{Eq:hat-M} 
\mathbf{\hat M}=\left(\begin{array}{cccc}
0&\xi^2\theta&\hbar&0 \\
-\xi^2\theta&0&0&\hbar \\
-\hbar&0&0&\xi^2\eta \\
0&-\hbar&-\xi^2\eta&0
\end{array}\right)
\end{equation}
The corresponding matrixes in commutative space are a $1\times 4$
column matrix $U$ with elements $U_1=x_1$, $U_2=x_2$, $U_3=p_1$
and $U_4=p_2$, a $4\times 1$ row matrix $U^T$ with elements
$U_i^T=U_i$, $(i=1,2,3,4)$, and a $4\times 4$ matrix $M$ with
elements
\begin{equation}
\label{Eq:Mij}
iM_{ij}=[U_i, U_j^T], (i, j=1,2,3,4).
\end{equation}
The matrix $M$ represents the undeformed Heisenberg-Weyl algebra,
which can be obtained by putting $\theta=\eta=0$ in the matrix
$\hat M$ (\ref{Eq:hat-M}),
\begin{equation}
\label{Eq:M} 
\mathbf{M}=\left(\begin{array}{cccc}
0&0&\hbar&0 \\
0&0&0&\hbar \\
-\hbar&0&0&0 \\
0&-\hbar&0&0
\end{array}\right)
\end{equation}

From Eq.~(\ref{Eq:hat-x-p}) it follows that $\hat U_i=R_{ik}U_k$,
$\hat U^T_j=\hat U_j==R_{jl}U_l=U^T_lR^T_{lj}$, and the deformed
Heisenberg-Weyl algebra is related to the undeformed
Heisenberg-Weyl algebra by a similarity transformation
\begin{equation}
\label{Eq:hat-M,M} 
\hat M_{ij}=R_{ik}M_{kl}R^T_{lj}
\end{equation}
 with a real matrix $R$
\begin{equation}
\label{Eq:matrix-R} 
\mathbf{R}=\left(\begin{array}{cccc} \xi&
0&0&-\frac{1}{2\hbar}\xi\theta \\
0&\xi&\frac{1}{2\hbar}\xi\theta&0 \\
0&\frac{1}{2\hbar}\xi\eta&\xi&0 \\
-\frac{1}{2\hbar}\xi\eta&0&0&\xi
\end{array}\right),
\end{equation}
It is worth noting that $R$ is not orthogonal matrix $RR^T\ne I$.
The first part of the un-equivalency theorem is proved.

Now we prove the second part of the un-equivalency theorem.
Eq.~(\ref{Eq:hat-x-p}) shows that if there is such a unitary
transformation, its elements should be real. That is, it should be
an orthogonal matrix $S$ with real elements $S_{ij}$,
$SS^T=S^TS=I$, and satisfies $S_{ik}\hat M_{kl} S^T_{lj}=M_{ij}$,
or $S_{ik}\hat M_{kj}=M_{ik} S_{kj}$. This is a system of 16
homogeneous linear equations for $S_{ij}$, $(i,j=1, 2, 3, 4)$. It
is divided into 4 closed sub-systems of 4 homogeneous linear
equations. Among them we consider a closed sub-system including
$S_{12}$, $S_{13}$, $S_{31}$ and $S_{34}$, which reads
\begin{subequations}
\begin{equation}
\label{Eq:4eqs-a}
\xi^2\theta S_{12}+\hbar S_{13}=-\hbar S_{31}
\end{equation}
\begin{equation}
\label{Eq:4eqs-b}%
 \hbar S_{12}+\xi^2\eta S_{13}=\hbar S_{34}
\end{equation}
\begin{equation}
\label{Eq:4eqs-c}%
\xi^2\theta S_{31}-\hbar S_{34}=-\hbar S_{12}
\end{equation}
\begin{equation}
\label{Eq:4eqs-d}%
\hbar S_{31}-\xi^2\eta S_{34}=-\hbar S_{13}
\end{equation}
\end{subequations}
(The situation for the rest elements is the same). The condition
of non-zero solutions of $S_{12}$, $S_{13}$, $S_{31}$ and $S_{34}$
is
\footnote {\; In Eq.~(\ref{Eq:non-zero}) dimensions of different
terms are different. If we define a $1\times 4$ column matrix
$\hat V$ with elements $\hat V_1=\hat x_1$, $\hat V_2=\hat x_2$,
$\hat V_3=\alpha\hat p_1$ and $\hat V_4=\alpha\hat p_2$, where
$\alpha$ is an auxiliary arbitrary non-zero constant with the
dimension $[mass]^{-1}[time]^1$. Thus $\hat V_i$ $(i=1, 2, 3, 4)$
have the same dimension $[lenth]^2$. Then in
Eq.~(\ref{Eq:non-zero}) dimensions of different terms are same.
The introduction of the arbitrary constant $\alpha$ does not
change the following conclusion.}
\begin{equation}
\label{Eq:non-zero} 
\xi^2\theta\eta=\pm \hbar(\theta+
\eta).
\end{equation}

In order to elucidate the physical meaning of
Eq.~(\ref{Eq:non-zero}), we consider conditions of guaranteeing
Bose-Einstein statistics in the case of both position-position and
momentum-momentum noncommuting in the context of non-relativistic
quantum mechanics. We start from the general construction of
deformed annihilation and creation operators $\hat a_i$ and $\hat
a_i^\dagger$ $(i=1,2)$ at the deformed level, which are related to
the deformed phase space variables $\hat x_i$ and $\hat p_i$. The
general form of $\hat a_i$ can be represented as $\hat
a_i=c_1(\hat x_i +ic_2\hat p_i)$, where constants $c_1$ and $c_2$
can be fixed as follows. The deformed annihilation and creation
operators $\hat a_i$ and $\hat a_i^\dagger$ should satisfy $[\hat
a_1,\hat a_1^\dagger]=[\hat a_2,\hat a_2^\dagger]=1$ (to keep the
physical meaning of $\hat a_i$ and $\hat a_i^\dagger$). From this
requirement and the deformed Heisenberg-Weyl algebra (\ref{Eq:xp})
it follows that $c_1=\sqrt{1/2c_2\hbar}$. When the state vector
space of identical bosons is constructed by generalizing
one-particle quantum mechanics, Bose-Einstein statistics should be
maintained at the deformed level described by $\hat a_i$, thus
operators $\hat a_1$ and $\hat a_2$ should be commuting. From
$[\hat a_i,\hat a_j]=0$ and the deformed Heisenberg-Weyl algebra
(\ref{Eq:xp}) it follows that
$ic_1^2\xi^2\epsilon_{ij}(\theta-c_2^2\eta)=0$, i.e.
$c_2=\sqrt{\theta/\eta}$. (The phases of $\theta$ and $\eta$ are
chosen so that $\theta/\eta>0$.) The general representations of
the deformed annihilation and creation operators $\hat a_i$ and
$\hat
a_i^\dagger$ are 
\begin{eqnarray}
\label{Eq:hat-a}
\hat a_i=\sqrt{\frac{1}{2\hbar}\sqrt{\frac{\eta}{\theta}}}\left
(\hat x_i +i\sqrt{\frac{\theta}{\eta}}\hat p_i\right),
\hat
a_i^\dagger=\sqrt{\frac{1}{2\hbar}\sqrt{\frac{\eta}{\theta}}}\left
(\hat x_i-i\sqrt{\frac{\theta}{\eta}}\hat p_i\right).
\end{eqnarray}
The structure of the deformed annihilation operator $\hat a_{i}$
in Eq.~(\ref{Eq:hat-a}) is determined by the deformed
Heisenberg-Weyl algebra (\ref{Eq:xp}) itself, independent of
dynamics. The special feature of a dynamical system is encoded in
the dependence of the factor $\eta/\theta$ on characteristic
parameters of the system under study.

In the limits $\theta,\eta\to 0$ and $\eta/\theta$ keeping finite,
the deformed annihilation operator $\hat a_i$ should reduce to the
undeformed annihilation operator $a_i$.
In commutative space in the context of non-relativistic quantum
mechanics the general form of the undeformed annihilation operator
$a_i$ can be represented as $a_i=d_1(x_i +id_2p_i)$. From $[a_1,
a_1^\dagger]=[a_2, a_2^\dagger]=1$ and the undeformed
Heisenberg-Weyl algebra (\ref{Eq:xp1}) it also follows that
$d_1=\sqrt{1/2d_2\hbar}$ with $d_2>0$. From Eq.~(\ref{Eq:xp1}) the
equation $[a_i,a_j]=0$ is automatically satisfied, thus there is
not constraint on the coefficient $d_2$. The general form of the
undeformed annihilation operator reads
\begin{equation}
\label{Eq:a} 
a_i=\sqrt{\frac{1}{2d_2\hbar}}\left (x_i +id_2 p_i\right).
\end{equation}
Like the situation of the deformed annihilation operator $\hat
a_{i}$, here the structure of $a_{i}$ is determined by the
undeformed Heisenberg-Weyl algebra (\ref{Eq:xp1}) itself,
independent of dynamics. The special feature of a dynamical system
is encoded in the dependence of the factor $d_2$ on characteristic
parameters of the system under study.
\footnote {\; For example, characteristic parameters of two
dimensional isotropic harmonic oscillator are the mass $\mu$, the
frequency $\omega$ and the constant $\hbar$. The dimension of
$d_2$ is $M^{-1}T$. The unique product of $\mu^{t_1}$,
$\omega^{t_2}$ and $\hbar^{t_3}$ possessing the dimension
$M^{-1}T$ is $\mu^{-1}\omega^{-1}$. Thus
$d_2=\kappa\mu^{-1}\omega^{-1}$, where $\kappa$ is a dimensionless
positive constant.}
If noncommutative quantum theory is a realistic physics, all
quantum phenomena should be reformulated at the deformed level.
This means that in the limits $\theta,\eta\to 0$ and $\eta/\theta$
keeping finite the deformed annihilation operator $\hat a_{i}$
should reduce to the undeformed one $a_{i}$. Comparing
Eq.~(\ref{Eq:hat-a}) and (\ref{Eq:a}), it follows that in the
limits $\theta,\eta\to 0$ and $\eta/\theta$ keeping finite the
factor $\eta/\theta$ reduces to a {\it positive} quantity:
\begin{equation}
\label{Eq:eta-theta}
\frac{\eta}{\theta}\to \frac{1}{d_2^2}>0.
\end{equation}
But from Eq.~(\ref{Eq:non-zero}), we obtain
$\eta/\theta=\pm\hbar/(\xi^2\theta\mp\hbar)$. This equation shows
that in the limits $\theta,\eta\to 0$ and $\eta/\theta$ keeping
finite, we have $\eta/\theta\to -1$, which contradicts
Eq.~(\ref{Eq:eta-theta}). We conclude that Eq.~(\ref{Eq:non-zero})
is un-physical. Thus the supposed orthogonal real matrix $S$
consistent with physical requirements does {\it not} exist. The
second part of the un-equivalency theorem is proved.

Now we consider the first condition about equivalency of the two
algebras. Eq.~(\ref{Eq:hat-x-p}) shows that the determinant
$\mathcal{R}$ of the transformation matrix $R$ between $(\hat
x_1,\hat x_2,\hat p_1,\hat p_2)$ and $(x_1,x_2,p_1,p_2)$ is
\begin{equation}
\label{Eq:det-R}
\mathcal{R}=\xi^4(1-\frac{\theta\eta}{4\hbar^2})^2.
\end{equation}
When $\theta\eta=4\hbar^2$, the matrix $R$ is singular. In this
case the inverse of $R$ does not exit. It means that the first
condition about equivalency of two algebra is not satisfied.

Summarizing the above results we conclude that for the case of
both position-position and momentum-momentum noncommuting the
deformed and the undeformed Heisenberg-Weyl algebra are not
equivalent.

For the case of only position-position noncommuting, $\eta=0$, the
transformation matrix $R$ between $(\hat x_1,\hat x_2,\hat
p_1,\hat p_2)$ and $(x_1,x_2,p_1,p_2)$ reduces to the matrix
$R_0$,
\begin{equation}
\label{Eq:matrix-R0} 
\mathbf{R_0}=\left(\begin{array}{cccc} 1&
0&0&-\frac{1}{2\hbar}\theta \\
0&1&\frac{1}{2\hbar}\theta&0 \\
0&0&1&0 \\
0&0&0&1
\end{array}\right).
\end{equation}
Its determinant $\mathcal{R}_0\equiv 1$, which is singular-free.
But in this case $R_0$ is not an orthogonal matrix either,
$R_0R_0^T\ne I$. Furthermore, in this case the supposed orthogonal
real matrix $S$ reduces to $S_0$. The closed sub-system of 4
homogeneous linear equations including $S_{0,12}$, $S_{0,13}$,
$S_{0,31}$ and $S_{0,34}$ has only zero solutions. The supposed
orthogonal real matrix $S_0$ does {\it not} exist, either. We
conclude that for the case of only position-position noncommuting
the deformed and the undeformed Heisenberg-Weyl algebras are also
not equivalent.

\vspace{0.4cm}

In this paper the extensively tacit understandings about
equivalency between the deformed Heisenberg-Weyl algebra in
noncommutative space and the undeformed Heisenberg-Weyl algebra in
commutative space is clarified. Unlike the claim in literature,  a
similarity transformation with an orthogonal real matrix which
transforms the deformed Heisenberg-Weyl algebra to the undeformed
one does not exist, even for the case of only position-position
noncommuting. Elucidation of the un-equivalency theorem between
two algebras has basic meaning both in theory and practice. The
deformed Heisenberg-Weyl algebra is the foundation of
noncommutative quantum theories. If  the deformed Heisenberg-Weyl
algebra were equivalent to the undeformed Heisenberg-Weyl algebra,
essentially new physics emerged from spatial noncommutativity
would not be expected. In practice, this theorem will inspire
efforts in searching for new effects of spatial noncommutativity
which is relevance to the real word, and hopefully shine some new
light on physical reality \cite{JZZ04b}.

\vspace{0.4cm}

This work has been supported by the National Natural Science
Foundation of China under the grant number 10575037 and by the
Shanghai Education Development Foundation.

\clearpage



\begin{thebibliography}{99}

\bibitem{DH}
A. Connes, M. R. Douglas, C. M. Hull,
JHEP {\bf 9802}, 008 (1998).

\bibitem{SW}
N. Seiberg and E. Witten,
JHEP {\bf 9909}:032 (1999).

\bibitem{DN}
M. R. Douglas, N. A. Nekrasov,
Rev. Mod. Phys. {\bf 73}, 977 (2001) and references there in.

\bibitem{CST}
M. Chaichian, M. M. Sheikh-Jabbari, A. Tureanu,
Phys. Rev. Lett. {\bf 86}, 2716 (2001).

\bibitem{GLR}
J. Gamboa, M. Loewe, J. C. Rojas,
Phys. Rev. {\bf D64}, 067901 (2001).

\bibitem{SS}
A. Smailagic, E. Spallucci,
Phys. Rev. {\bf D65}, 107701 (2002).

\bibitem{Acat}
C. Acatrinei,
JHEP {\bf 0109}, 007 (2001).

\bibitem{HK}
P-M. Ho, H-C. Kao,
Phys. Rev. Lett. {\bf 88}, 151602 (2002).

\bibitem{NP}
V. P. Nair, A. P. Polychronakos,
Phys. Lett. {\bf B505}, 267 (2001). 

\bibitem{JM}
L. Jonke, S. Meljanac,
Euro. Phys. J. {\bf C29}, 433 (2003). 

\bibitem{CDPST00}
M. Chaichian, A. Demichev, P. Presnajder, M.M. Sheikh-Jabbari, A.
Tureanu,
Phys. Lett. {\bf B527}, 149 (2002).

\bibitem{CDPST01}
M. Chaichian, A. Demichev, P. Presnajder, M.M. Sheikh-Jabbari, A.
Tureanu,
Nucl. Phys. {\bf B611}, 383 (2001).

\bibitem{Suss}
L. Susskind ,
{\bf hep-th/0101029}.

\bibitem{FGLMR}
H. Falomir, J. Gamboa, M. Loewe, F. Mendez, J.C. Rojas,
Phy. Rev. {\bf D66}, 045018 (2002).

\bibitem{JZZ04a}
Jian-zu Zhang,
Phys. Lett. {\bf B584}, 204 (2004); 
Jian-zu Zhang,
Phys. Rev. Lett. {\bf 93}, 043002 (2004); 
Jian-zu Zhang,
Phys. Lett. {\bf B597}, 362 (2004); 
Qi-Jun Yin, Jian-zu Zhang,
Phys. Lett. {\bf B613}, 91 (2005). 

\bibitem{JZZ04b}
Jian-zu Zhang,
Phys. Rev. Lett. {\bf 93}, 043002 (2004). 

\end{thebibliography}
\end{document}